\documentclass[useAMS,usenatbib]{mn2e}
\usepackage{graphicx,amssymb, epstopdf}
\newcommand{\vc}[1]{\bmath{#1}}

\newcommand{\beq}{\begin{equation}}
\newcommand{\eeq}{\end{equation}}
\newcommand{\bea}{\begin{eqnarray}}
\newcommand{\eea}{\end{eqnarray}}

\newcommand{\figref}[1]{Fig.~\ref{#1}}
\newcommand{\Secref}[1]{Section~\ref{#1}}

\newcommand{\eqref}[1]{equation~(\ref{#1})}

\newcommand{\dd}{\partial}
\newcommand{\did}{{\rm d}}

\newcommand{\vthi}{v_{\rmn{th}i}}
\newcommand{\vthe}{v_{\rmn{th}e}}

\newcommand{\para}{\parallel}

\newcommand{\nesim}{n_{e,{\rm sim}}}
\newcommand{\nez}{n_{e0}}
\newcommand{\Tproj}{T_{e,{\rm proj}}}
\newcommand{\Tsim}{T_{e,{\rm sim}}}

\title[Weak shocks in cool-core galaxy clusters]{Propagation of weak shocks in cool-core galaxy clusters in two-temperature magnetohydrodynamics with anisotropic thermal conduction}
\author[S. Komarov,  C. Reynolds and E. Churazov]{
S. Komarov$^{1}$\thanks{E-mail: skomarov@ast.cam.ac.uk}, C. Reynolds$^{1}$ and E. Churazov$^{2,3}$\\
$^{1}$Institute of Astronomy, University of Cambridge, Madingley Road, Cambridge CB3 0HA, United Kingdom\\
$^{2}$Max Planck Institute for Astrophysics, Karl-Schwarzschild-Strasse 1, 
85741 Garching, Germany\\
$^{3}$Space Research Institute (IKI), Profsouznaya 84/32, Moscow 117997, Russia
} 

\begin{document}
\maketitle
\label{firstpage}

\begin{abstract}
We investigate how different magnetohydrodynamic models of propagation of a weak (Mach number $\sim1.2$) 
shock in the core of a galaxy cluster affect its observational appearance, using the Perseus cluster 
as our fiducial model. In particular, we study how thermal conduction, both isotropic and anisotropic, 
and ion-electron temperature equilibration modify a weak shock.
Strong thermal conduction is expected to produce an electron temperature precursor. 
Less prominent pressure and density precursors are formed as well. A longer equilibration 
time largely reduces the density precursor, 
but does not change the electron temperature precursor much. When thermal conduction becomes 
anisotropic, the intracluster magnetic field imprints its characteristic spatial scale on the 
distortions of the shock induced by heat fluxes.

\end{abstract}


\section{Introduction}
\label{sec:intro}

Active galactic nuclei (AGN) in the cool cores of galaxy clusters inject an enormous amount 
of energy into the intracluster medium (ICM) by launching powerful relativistic jets. This 
process is often thought to maintain the thermodynamic stability of cluster cores by offsetting 
radiative cooling of the dense plasma (e.g., \citealt{Chur2000,Rusz2004,Fabian2005,
Zhur2016}; for reviews, see \citealt{Fabian_rev,Werner_rev}). AGN jets produce a variety of physical structures: 
they inflate bubbles of relativistic plasma that rise in the cluster atmosphere potentially driving 
turbulence, launch weak shocks and sound waves, as well as inject cosmic rays into the ICM. Whether one of these processes 
dominates the heating of cool cores has been a matter of active debate for two decades 
(see \citealt{Chur2000,Reynolds2002,Dennis2005,Zhur2014,Zhur2016,Forman2017} for turbulent heating; 
\citealt{Rusz2004,Fabian2005,Fujita2005,Shabala2009,Sternberg2009,Fabian2017} for 
heating by sound waves; for cosmic-ray heating, see, e.g., \citealt{Fulai2008}). In the second scenario, sound waves 
propagate through the cores and gradually dissipate their energy via thermal conduction. 
The efficiency of thermal conduction sets whether sound waves are able to deploy the energy 
over a significant fraction of the core volume before they decay. 
%Thermal conduction in the ICM also plays a key role in the development of buoyancy 
%instabilities, such as the heat-flux driven buoyancy instability in the cool cores (REFS) 
%and magnetothermal instability on the cluster outskirts (REFS). 
While magnetic fields in the ICM have typical magnitudes 1--10 $\mu$G \citep[e.g.,][]{Feretti2012}, they force 
the plasma particles to spiral around magnetic field lines with microscopic Larmor radii, 
so that particle diffusion across field lines is almost fully inhibited, and only parallel transport is allowed. 
The accumulating observational evidence based on analyzing temperature variations and radial 
profiles in galaxy clusters indicates that thermal conduction in the bulk of the ICM should be suppressed 
by factors of up to several tens (see, e.g., \citealt{Mark2003,Fulai2018} for
estimates of large-scale suppression not associated with peculiar structures in the ICM where 
thermal isolation by magnetic-field draping may be dominant), 
likely beyond simple magnetohydrodynamic 
estimates of suppression due to tangled magnetic fields (factors $\sim3$--$5$; see 
\citealt{Narayan2001}). 
Current theoretical evidence and particle-in-cell plasma simulations provide rather modest  
suppression factors of parallel thermal conductivity up to 
$\sim10$ \citep{Komarov2016,RC2016,RC2017,Komarov2018}. While the overall suppression 
can already be explained by a combination of effects resulting from the micro-scale plasma kinetics 
(that reduces parallel conductivity) and 
tangled magnetic field topology (that slows down the large-scale transport of heat), the 
relative contribution of parallel suppression is mainly unclear. It is the local parallel 
thermal conductivity that is relevant for the dissipation of sound waves. Therefore, it is helpful to look for 
signatures of \textit{parallel} suppression while studying various structures in the ICM, both in 
observations and simulations. 

Deep {\it Chandra} X-ray observations of the cool cores in the Perseus cluster and in the giant elliptical galaxy M87 
reveal remarkably regular spherical weak shocks (for the Perseus cluster, 
see \citealt{Fabian2006}; for M87, \citealt{Forman2007}). These shocks have very low Mach 
numbers $\approx 1.2$ and, therefore, propagate 
almost at the speed of sound. The ion and electron Coulomb mean free paths in their vicinity 
are $\sim 0.1$ kpc, below the resolution of {\it Chandra}. If electron 
thermal conduction across very weak shocks is not suppressed by kinetic instabilities, it can 
appreciably reduce the sharpness of temperature and even density jumps for the parameters 
characteristic of the Perseus and M87 cores. In this case, thermal electrons diffuse rapidly in 
front of a shock, forming a temperature and, to a lesser extent, pressure and density precursors.  
Also, because heat is transferred strictly along magnetic field lines, their topology may be 
imprinted on the shock producing temperature/density inhomogeneities along the front. 
In modelling such effects, it is important to note that the electron-ion energy equilibration 
time scale is three orders of magnitude (the proton-to-electron mass ratio) longer than 
the time between electron collisions. This means that while the jump of the electron temperature 
becomes smoothed, that of the ion temperature and plasma density (electron and ion densities 
are the same due to plasma quasineutrality) may remain much sharper.

In this paper, we perform a number of two-temperature magnetohydrodynamics
simulations of weak shocks including the effects of anisotropic thermal conduction and long 
electron-ion equilibration time. We use the weak shock $\approx25$ kpc north-east of the 
Perseus center as our fiducial 
model. The weak shock in M87 has similar properties, however, the {\it Chandra} observations 
of the Perseus core we use for comparison with our simulations are of much higher quality due 
to the higher brightness of the Perseus cluster.  
%mainly because of the fact that the plasma density and electron mean free path stay 
%roughly the same within the shock radius in Perseus, as opposed to M87, where the density increases 
%by an order of magnitude towards the center \citep[e.g.,][]{Forman2007}, thus reducing 
%the strength of the effects we aim to study . 
Hydrodynamic modelling of the Perseus weak shock with 
thermal conduction has been done by a number of authors \citep{Graham2008,TC2018}. 
We attempt to expand on 
their work by adding magnetic fields and a finite electron-ion temperature equilibration time. 
The paper is organized as follows. In 
\Secref{sec:model}, we describe the physical model and numerical set-up, then in 
\Secref{sec:results}, we present results of our numerical simulations and compare 
them with observations of the Perseus weak shock. \Secref{sec:disc} discusses the validity 
of our assumptions and potential significance of the results for current and future X-ray 
observations. Finally, \Secref{sec:concl} summarizes our findings.

\section{Model}

\label{sec:model}

\subsection{Equations solved}

We solve the equations of two-temperature magnetohydrodynamics joined by a heat-flux term that 
describes anisotropic electron thermal conduction in a magnetic field 
%%\footnote{For weak magnetic fields typical of galaxy clusters, their dynamic effects on the propagation of a shock are small, while their geometry is almost unaffected when the shock is weak} 
and a term that describes electron-ion equilibration as follows

\bea
\lefteqn{
\frac{\partial \rho}{\partial t} + \nabla\cdot (\rho \vc{v})=0,
\label{eq:mass_cons}
}&&\\
\lefteqn{
\frac{\partial \rho \vc{v}}{\partial t} + \nabla\cdot 
\left (\rho \vc{v}\vc{v} - \frac{\vc{B} \vc{B}}{4\pi} \right )
+ \nabla p = 0,
}&&\\
\lefteqn{
\frac{\dd E}{\dd t} + \nabla\cdot \left [ \vc{v} (E+p) - \frac{\vc{B}(\vc{v}\cdot\vc{B}) }{4\pi} \right ] = - \nabla\cdot Q,
}&&\\
\lefteqn{
\frac{\dd \vc{B}}{\dd t} + \nabla \cdot (\vc{B} \vc{v} - \vc{v} \vc{B}) = 0,
\label{eq:induct}
}&&\\
\lefteqn{
\frac{\dd s_e}{\dd t} + \nabla\cdot (s_e \vc{v}) = 
\left [ \frac{{\rm d} s_e}{{\rm d}t} \right ]_{ei} + 
\left [ \frac{{\rm d} s_e}{{\rm d}t} \right ]_{ee},
\label{eq:se}
}
\eea
where
\bea
p&=&p_i + p_e + \frac{B^2}{8\pi},\\
E &=& \frac{\rho v^2}{2} + \frac{p_i+p_e}{\gamma-1} + \frac{B^2}{8\pi},\\
s_e &=& \rho \log (p_e / \rho^{\gamma}),\\
Q &=& -\kappa_{\para} \vc{b}\vc{b}:\nabla T_e,
\label{eq:hf}
\eea
where $\rho$ is the mass density (taken to be the ion mass density), $\vc{v}$ the fluid 
velocity (assumed same for both species), $p_i$ and $p_e$ are the ion and electron 
pressures, $\vc{B}$ the magnetic field, $s_e$ the electron entropy, and $Q$ the heat flux along the magnetic field 
lines whose direction is set by unit vectors $\vc{b}$. Both electron and ion components 
are described by an ideal equation of state $p_{e/i}=n_{e/i} T_{e/i}$ (temperature is in energy units), where $n_e=n_i=
\rho/m_i$ are the particle number densities (equal by plasma quasineutrality), and 
adiabatic index $\gamma=5/3$. For the parallel thermal conductivity, we 
use the Spitzer value \citep{Spitzer1962} $\kappa_{\para}\approx 0.9 n_e \lambda_e 
v_{{\rm th}e}$, where $n_e$ is the electron density, $v_{{\rm th}e}=(2T_e/m_e)^{1/2}$ 
is the electron thermal speed, and 
\beq
\label{eq:mfp}
\lambda_e \approx 0.1~{\rm kpc} 
\left ( \frac{T_e}{4~{\rm keV}} \right )^2  
\left ( \frac{n_e}{4 \times 10^{-2}~{\rm cm}^{-3}}  \right )^{-1} 
\eeq
is the Coulomb electron mean free path in a hydrogen plasma expressed using 
characteristic parameters of the Perseus core in the vicinity of the observed weak shock. 
The two terms on the right-hand 
side of \eqref{eq:se} describe dissipation due to electron-ion and electron-electron 
collisions.

Numerically, the dissipation terms are treated separately at the end of each time 
(sub-)step. First, the system of conservation laws for ($\rho$, $\rho \vc{v}$, $\vc{B}$, $E$, 
$s_e$) described by equations~(\ref{eq:mass_cons})--(\ref{eq:se}) with the RHS set to 
zero is evolved by a second-order van Leer integrator using the HLLD flux \citep{hlld2005} 
modified to include electron pressure. The magnetic field is integrated via the constrained 
transport technique \citep{vanleer2009}.  Second, electron thermal conduction is added by 
solving anisotropic heat transfer equation
\beq
\frac{\rho}{(\gamma-1) m_i}\frac{\dd T_e}{\dd t} = - \nabla\cdot Q
\eeq
using Runge-Kutta-Legendre super-time-stepping \citep{Meyer2012,Meyer2014} 
with a monotonized 
central limiter applied to transverse temperature gradients to avoid temperature 
oscillations \citep{Sharma2007,Sharma2011}. Finally, electron-ion equilibration is accounted for by implicit 
integration (this might be a numerically stiff problem when the equilibration is fast) 
of equation
\beq
\frac{\dd (T_e-T_i)}{\dd t} = - \nu_{ei} (T_e-T_i),
\eeq
where
\bea
\nonumber
\nu_{ei} &\approx& 4 \frac{m_e}{m_p} \frac{v_{{\rm th}e}}{\lambda_e} \\
&\approx& 5 \times 10^{-16}~{\rm s}^{-1}
\left ( \frac{T_e}{4~{\rm keV}} \right )^{-3/2}  
\left ( \frac{n_e}{4 \times 10^{-2}~{\rm cm}^{-3}}  \right )
\eea
is the thermal equilibration rate \citep{NRL}. The new electron temperature is then used to 
update the electron entropy $s_e$ (which corresponds to 
adding the two dissipative terms on the RHS of equation \ref{eq:se}) 
and total energy density $E$ (only due to thermal conduction).

We also perform a series of single-fluid hydrodynamic runs with or without thermal 
conduction, where the same system of equations is solved with $p=2p_i=2p_e=2 
n_e T_e$ excluding the induction equation~(\ref{eq:induct}) and \eqref{eq:se} for the 
electron entropy, as well as runs with isotropic 
thermal conduction, where instead of \eqref{eq:hf}, the heat flux is
$Q = -\kappa_{\para} \nabla T_e$.

\subsection{Numerical setup}

\label{sec:setup}

As we are not concerned about the distribution of energy released during an AGN 
outburst between the radio bubble and the weak shock, it is sufficient to produce 
a spherical blast wave by the simplest suitable initial condition. The only requirement 
is that the shock needs to have Mach number $M\approx 1.2$ as it reaches radius 
$r\approx 15$ kpc to match observations of the weak shock in the Perseus cluster. 
Therefore, for the initial conditions, we 
produce an overpressurized (by a factor of $\sim 100$) region within $r<1.5$ kpc, 
large enough to avoid pixelation. The weak shock produced in the outburst travels outward unimpeded, 
while the hot gas shell formed at the center is of no relevance for our problem. 
Following the outburst, we observe how the shock is modified by thermal 
conduction and electron-ion equilibration as it propagates to $r=15$ kpc. 
Our 3D numerical box has size $L=40$ kpc and is centered around the shock. We choose 
to use Cartesian coordinates because of the easier implementation of runs with 
tangled magnetic fields. 

The observed part of the spherical weak shock in the Perseus cluster is located 
$\approx 25$ kpc away from the cluster center in the north-east direction (its geometrical 
center is $\approx10$ kpc off the cluster center).  In the 
vicinity of the shock, the ICM has temperature 
$T_0\approx4$ KeV and the radial temperature profile is flat \citep[e.g.,][]{Graham2008,Zhur2015}. 
The ICM density in this region is $n_0\approx0.04~{\rm cm}^{-3}$. In our simulations, 
we choose to use uniform background temperature and density for several reasons. 
First, both radial profiles within the shock sphere are relatively flat \citep[e.g.,][]{Zhur2015}. 
Second, there exists uncertainty about the initial density and temperature distribution at 
the moment of the AGN outburst that launched the shock, and it is likely that 
the initial distribution was different from the one currently observed. 
More importantly for our particular problem, the structures we are interested in, 
i.e., distortions of the shock due to thermal conduction, develop quickly (compared 
to the time it takes for the shock to reach its current position) and locally, 
within $\sim 5$ kpc from the shock front. Therefore, it is sufficient 
to initialize the background density and temperature to those currently observed near the 
shock, as long as the simulated shock has the right Mach number $M\approx1.2$ 
at radius $r\approx15$ kpc.  
%We assume a uniform background ICM temperature $T=4$ KeV inside the 
%shock sphere due to the relative flatness of the Perseus radial temperature profile 
%in this region (based on the deprojected temperature profiles in, e.g., \citealt{Graham2008,Zhur2015}). 
%For the background density profile, we use a $\beta$-model with central density 
%$n_c=0.05~{\rm cm}^{-3}$, core radius $r_c=29$ kpc, and $\beta=0.53$ 
%\citep{Zhur2015,Zhur2016}.

There is currently a lot of uncertainty about the structure of magnetic field 
in the ICM, mainly due to the complicated physics of turbulent magnetized 
plasmas at high ratios of thermal to magnetic pressures and, on the other 
hand, limited possibilities to infer a 3D spectrum of magnetic fluctuations 
from radio observations \citep[e.g.,][]{Carilli2002,Govoni2004,Feretti2012,Vogt2005,
Bonafede2010,Kuchar2011}. The magnetic spectrum is likely anisotropic, 
meaning the length scale of change of the field along itself is different (normally 
larger) than that transverse to itself \citep[e.g.,][]{Schek2006turb}. The parallel correlation length 
is expected to be set by large-scale hydrodynamic motions (on those scales 
the field is too weak to have any dynamic effect on the fluid). But the perpendicular 
correlation scale depends on the intricacies of plasma turbulence and ranges 
from Larmor to injection scales depending on the numerical experiment 
and its interpretation
(for MHD/Braginskii turbulence, see 
\citealt{Haugen2004,Schek2004,Maron2004,Cho2009,Beres2009,Beres2012,Lima2014,Onge2020}; 
for kinetic turbulence, see \citealt{Rincon2016,Onge2018}).
Rotation measure 
(meaning rotation of the polarization plane of synchrotron radiation as it travels through 
a magnetic field) observations of some clusters and groups of galaxies give 
estimates of the perpendicular scale of order several kpc 
\citep{Vogt2005,Bonafede2010,Kuchar2011}. In any case, 
for the problem of anisotropic thermal conduction in a static (on time scales 
of shock propagation) tangled magnetic field, the more relevant correlation 
length is parallel. This is because heat diffusion is not affected by field 
reversals: diffusion operates in layers of antiparallel field lines the same way 
as in a uniform field. Then for qualitative purposes, it is sufficient to use a 
statistically isotropic field on scales characteristic of gas motions induced by 
AGN activity or gas sloshing in the Perseus cluster. These scales likely range from 
a few to several tens kpc corresponding to the size of structures (radio bubbles, 
filaments, sloshing-induced spiral patterns) seen in {\it Chandra} X-ray images 
\citep{Fabian2006}. For this reason, we model the magnetic field simply as a random isotropic 
solenoidal 3D field with coherence length $l_B\approx10$ kpc. The field is set to be very weak, so 
it does not have a dynamic feedback on the plasma. Such coherence length is 
somewhat larger than the width of conductive weak shocks in our simulations ($\sim5$~kpc). 
%A much smaller length is unjustified observationally due to the larger typical scales of AGN-produced 
%structures; it allows one to average over directions of the small-scale magnetic field 
%leading to a constant equivalent reduction of isotropic conduction by a factor 
%of order $1/3$.

Thermal conduction along the magnetic field lines operates via heat flux (\ref{eq:hf}) with 
electron mean free path (\ref{eq:mfp}). 
To reduce the unnecessary computational load, we turn off thermal conduction in the central 
6-kpc region that quickly becomes dominated by the hot shell. Initially, the ion and electron 
temperatures are set equal.

\section{Results}

\label{sec:results}

\subsection{Isotropic thermal conduction}

\begin{figure}
\centering
\includegraphics[width=0.9\columnwidth]{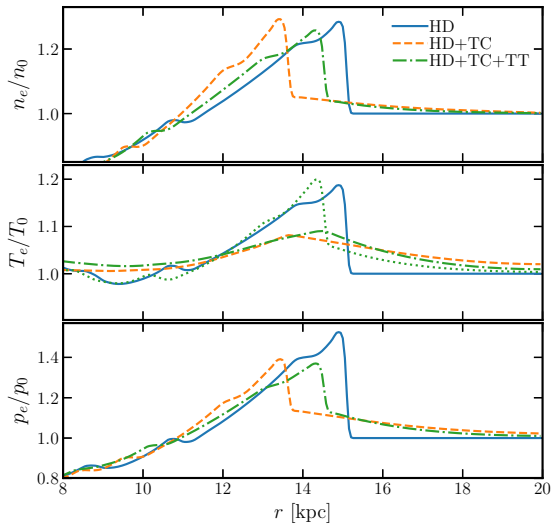}
\caption{Profiles of electron density (upper panel), temperature (middle panel), 
and pressure (lower panel) in 
3 different models: single-temperature hydrodynamics without (solid line) and with 
(dashed) thermal conduction and two-temperature hydrodynamics with conduction (dash-dotted). 
The dotted line shows the ion temperature in the third model. 
Both profiles are normalized to the pre-shock values. 
The $x$-axis shows the distance from the center of the outburst. }
\label{fig:1dsim}
\end{figure}

First, we analyze 1D hydrodynamic runs with a thermal conductivity at full Spitzer level. This 
models a configuration in which the magnetic field is radial and does not impede 
conduction. \figref{fig:1dsim} shows the effects of thermal conduction and 
a finite electron-ion equilibration time relative to pure single-fluid hydrodynamics. 
In pure hydrodynamics (solid line), the density and temperature jumps agree with the 
Rankine-Hugoniot relations for a shock with Mach number $M\approx 1.18$:
\bea
\frac{n_2}{n_1} &=& \frac{ \gamma+1 }{ \gamma-1 + 2/M^2 } \approx 1.27,\\
\frac{T_2}{T_1} &=& \frac{ (2\gamma M^2 - (\gamma -1) ) ((\gamma-1) M^2 + 2)}
									    { (\gamma+1)^2 M^2 } \approx 1.18.
\eea

Activating thermal conduction leads to formation of a temperature and, 
to a lesser extent, pressure and density precursors (dashed line in \figref{fig:1dsim}). Its 
length is consistent with the simple estimate of the distance $l_{\rm diff}$ that 
the electrons are able to diffuse in front of the shock based on equating the 
diffusion time and the time it takes for the shock to travel the diffusion length. 
The former is $t_{\rm diff} \sim l_{\rm diff}^2/ (\lambda_e \vthe)$, 
the latter $t_{\rm sh} \sim l_{\rm diff} / (M \vthi)$. 
Then $l_{\rm diff} \sim (m_i/m_e)^{1/2} \lambda_e$. 
Recalling that $\lambda_e\approx 0.1$ kpc (equation~\ref{eq:mfp}), we get 
$l_{\rm diff} \sim 4$ kpc, in agreement with \figref{fig:1dsim}. 
In comparison with ideal hydrodynamics, the temperature jump becomes 
smaller due to thermal conduction, and because the gas becomes 
closer to isothermal, the shock starts to slightly lag behind.

Enabling different temperatures (dash-dotted line in \figref{fig:1dsim}) 
for the electrons and ions leads to little changes of the electron temperature profile, 
but the density profile becomes much closer to the hydrodynamic case. 
This happens because the electrons diffusing ahead of the shock do not have 
sufficient time to heat the ions (the ion temperature is shown in the lower 
panel of \figref{fig:1dsim} by the dotted line). Because now the ions are 
mostly adiabatic, the speed of sound rises relative to the model with instant 
equilibration, and, therefore, the shock lags less.

Our single-temperature models produce weak 
shock profiles qualitatively similar to \cite{Graham2008} (see their Fig.~7; note that we do not match 
the shock front locations for different models in \figref{fig:1dsim}). Even though these authors use a more 
elaborate model of the outburst and introduce a smooth background density gradient, 
the weak shock itself is not sensitive to the 
initial conditions, and its profile is mainly determined by the intensity of thermal 
conduction across the shock. Our two-temperature model, however, is significantly different. 
While \cite{Graham2008} approximate the decoupling of electrons and ions by 
doubling the effective thermal conductivity, we solve a separate equation for the 
electron temperature, which is coupled to the ion temperature by a collision operator. 
This leads to noticeably different density and temperature profiles: namely, the ions 
tend to preserve their density and temperature jumps much better compared to 
the single-fluid model with conduction, while the electron temperature behaves 
the same as the gas temperature in the single-fluid model (compare this with the blue 
dotted line in Fig.~7 of \citealt{Graham2008}).

\begin{figure}
\centering
\begin{minipage}{0.95\columnwidth}
  \centering
  \includegraphics[width=0.85\columnwidth]{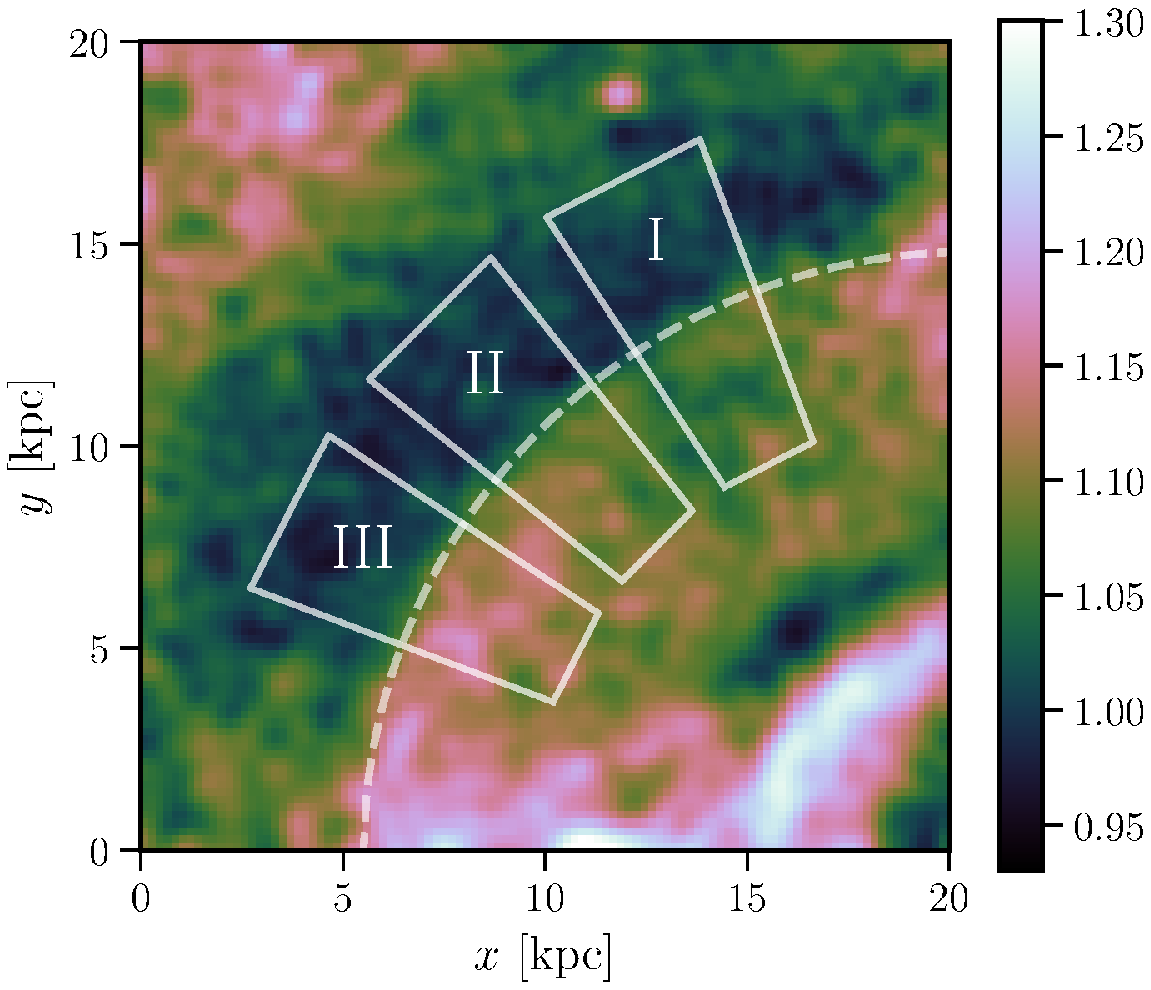}
  \caption{A {\it Chandra} X-ray image of the weak shock $\approx 10$ kpc 
  north-east of the center of the Perseus cluster. The image shows the X-ray surface 
  brightness divided by 
  a smooth $\beta$-model with parameters $\beta=0.53$ and $r_c=29$ kpc and 
  convolved with a Gaussian filter of size $\sigma\approx 1''\approx 0.4$ kpc. 
  The surface brightness is normalized to its average value in the pre-shock region.
  The superimposed sectors are the areas used to produce surface brightness 
  profiles in \figref{fig:sb1d}.}
  \label{fig:pers_sect}
  \vfill
  \includegraphics[width=0.88\columnwidth]{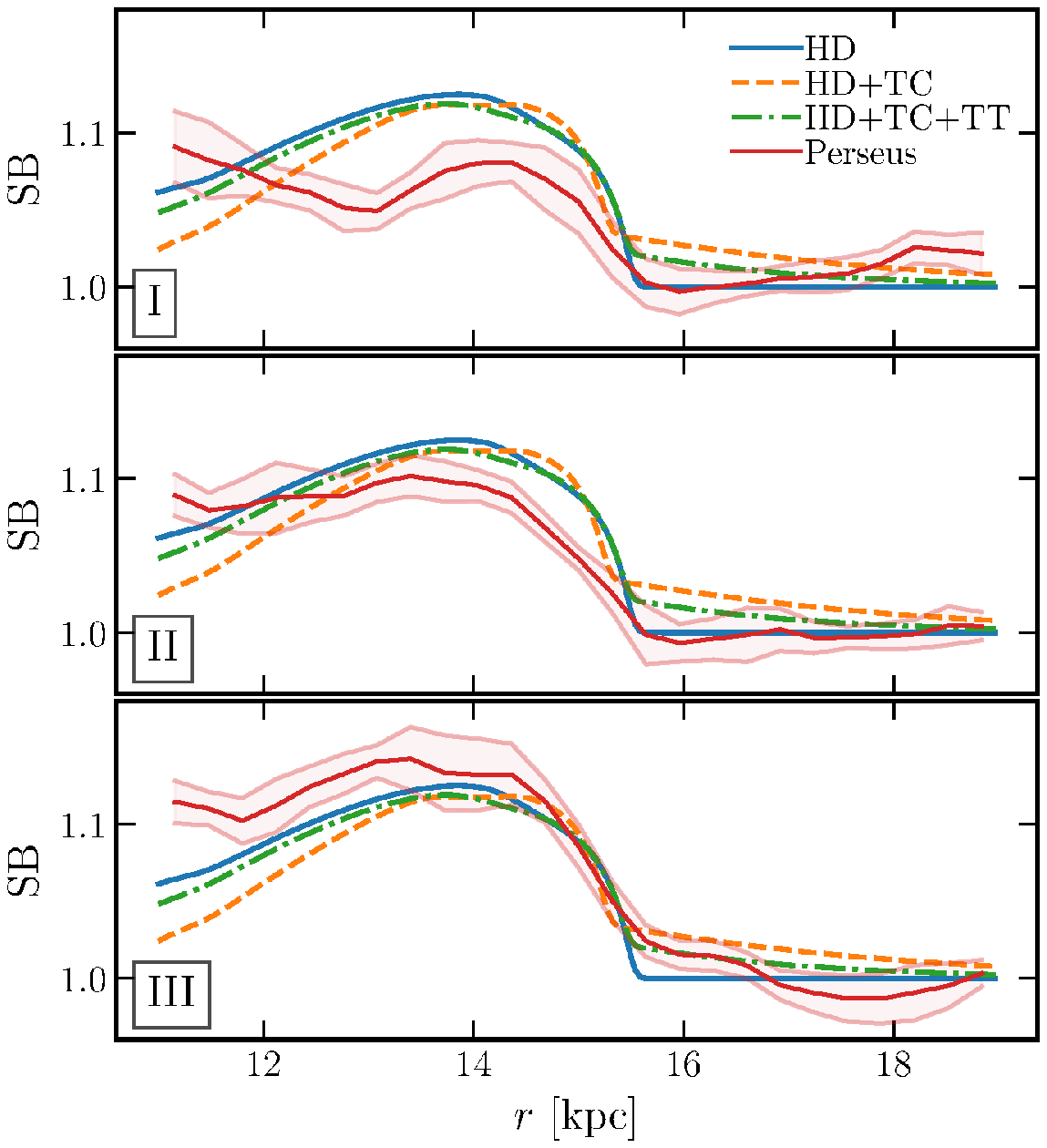}
  \caption{Surface brightness profiles calculated from the 1D simulations 
  (same as in \figref{fig:1dsim}) using \eqref{eq:SB}
  compared with those extracted from the Perseus X-ray image in \figref{fig:pers_sect}) 
  (red lines). The surface brightness is normalized to the pre-shock region.
  The $x$-axis is the distance from the geometrical center of the shock sphere.
  The error bars indicate the rms fluctuations of surface brightness in the 25 
  bins used for averaging over the sectors in \figref{fig:pers_sect}. 
  The shock locations in the different runs have been matched to facilitate the comparison.}
  \label{fig:sb1d}
\end{minipage}

\end{figure} 

We can use the simulated density and temperature profiles in combination 
with large-scale radial profiles observed in the Perseus cluster to produce 
mock surface brightness maps and compare them with X-ray observations
of the weak shock in the Perseus cluster. 
Given its spatial resolution, the most appropriate comparison data are from the {\it Chandra} Advanced Camera for Imaging Spectroscopy (ACIS).  We obtained all ACIS observations of the core of Perseus that did not employ the diffraction gratings, totalling 900ks of good data spread across 21 separate observations (ObsIDs).  Each of separate ObsIDs were reprocessed up to the calibration version 4.8.1 using CIAOv4.10, and then were merged into a single events file from which we extract a 0.7--4keV band image. 
We analyze a $40\times40$ kpc region of the Perseus core that contains the weak shock. 
The final image, shown in \figref{fig:pers_sect}, 
has been divided by an appropriate $\beta$-model with central density $n_c=0.05~{\rm cm}^{-3}$, 
core radius $r_c=29$ kpc, and $\beta=0.53$ \citep{Zhur2015} and smoothed with a 
Gaussian filter of size $\sigma\approx1''\approx 0.4$ kpc. Its bottom right corner 
coincides with the geometrical center of the shock sphere. In our energy range, the 
emissivity of an optically thin plasma is almost independent of temperature 
observed by the ACIS-S detector on {\it Chandra} \citep[e.g.,][]{Forman2007,Zhur2015}. Therefore, 
surface brightness is simply an integral of the electron density squared along 
the line of sight. 

In order to model the observed scaled X-ray surface brightness by integrating the emissivity 
along the line of sight, we need to know the 
large-scale ICM density distribution. In our simulations, we have used a uniform density 
and temperature background (see Section~\ref{sec:setup}). 
This is justified because the $\beta$-model in the vicinity of the shock 
(its width including the precursor is $\lesssim 10$~kpc) is smooth. Then we can 
factorize the large-scale electron density as $n_e(x,y,z)=\nez (r) \times \nesim (r_s)$, 
where $\nez (r) = n_c [1+(r/r_c)^2]^{-3/2 \beta}$ is the smooth background 
$\beta$-model and $\nesim (r_s)$ is the simulated electron density (shown in the upper panel 
of \figref{fig:1dsim}). 
The former is a function of the distance from the center of the cluster
$r=(x^2 + y^2 + z^2)^{1/2}$, where $x$, $y$ are the sky coordinates 
relative to the cluster center and $z$ is the distance along the line of sight. 
The latter is a function of the distance from the center of the shock sphere 
($x_{0s},y_{0s}$): $r_s=((x-x_{0s})^2 + (y-y_{0s})^2 + z^2)^{1/2} $. 
Then the mock surface brightness divided by the $\beta$-model is
\beq
{\rm SB}(x,y) = \frac{ \int_0^{\infty}  [\nesim(x,y,z) \times  \nez(x,y,z) ] ^2  \did z }
						 {  \int_0^{\infty} [ \nez(x,y,z) ]^2  \did z}.
\label{eq:SB}
\eeq
In \figref{fig:pers_sect}, the cluster center is located at $x\approx26$ kpc, $y\approx-8$ kpc.

In \figref{fig:sb1d}, we compare the mock surface brightness profiles for 
the three 1D models of the weak shock with surface brightness profiles 
obtained from the X-ray data by averaging over the three different sectors 
in \figref{fig:pers_sect}. The locations of the simulated shocks have been 
matched to facilitate the comparison. 
While it is rather difficult to distinguish between 
the pure hydrodynamic model and the one with both thermal conduction 
and ion-electron equilibration enabled, we see no indication of 
a conductive precursor, except for the third region, where variations 
of surface brightness before the shock preclude simple assessment. 
Deviations of the electron density from the $\beta$-model manifested 
in the significant variation of the observed profiles in the three chosen regions, 
particularly in the upstream region where precursors could be produced, 
also make further matching with our models problematic.

\begin{figure}
\centering
\includegraphics[width=0.9\columnwidth]{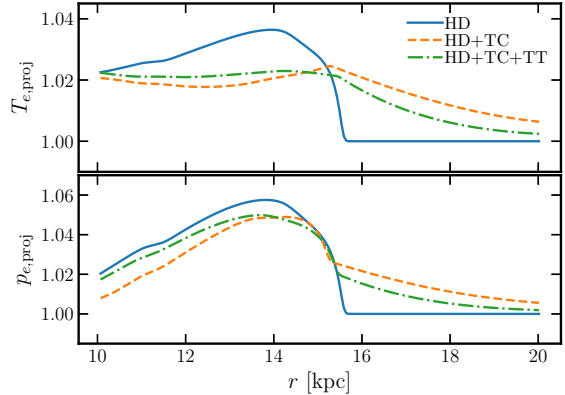}
\caption{Projected temperature (upper panel) and electron pressure 
(lower panel) profiles calculated from the 1D runs 
(same as in \figref{fig:1dsim}) using \eqref{eq:tproj} and \eqref{eq:sz}. 
Both quantities are normalized to the pre-shock values. 
The shock locations in the different runs have been matched.}  
\label{fig:tproj1d}
\end{figure} 

In addition, we can calculate the projected electron temperature $\Tproj (x,y)$ 
using the simulated data $\Tsim (r_s)$ (shown in the lower panel 
of \figref{fig:tproj1d}) by weighting the temperature with the emissivity and 
dividing it by the smooth background as follows
\bea
\nonumber
\Tproj &=& \frac{  \int_0^{\infty}  (\Tsim  \times T_{e0} )  ( \nesim  \times  \nez) ^2  \did z }
					  {  \int_0^{\infty} ( \nesim \times \nez )^2  \did z} \\
		&\times& \left [ \frac{ \int_0^{\infty}  T_{e0} \nez  ^2  \did z }
					  				     { \int_0^{\infty}  \nez ^2  \did z} \right ]^{-1} ,
\label{eq:tproj}
\eea
where
\beq
T_{e0} = T_{ec}  \frac{1. + (r/r_c)^{3.5}}{1 + 0.4(r/r_c)^{3.6}} 
\eeq
is the background temperature profile with $T_{ec}=4$ keV and $r_c=58$ kpc 
taken from \cite{Zhur2015}. We have omitted the coordinate 
$(x,y,z)$-dependence in \eqref{eq:tproj} for brevity. We have factorized the 
large-scale temperature analogously to the density. The projected temperature 
profiles as functions of the distance from the center of the outburst are 
demonstrated in the upper panel of \figref{fig:tproj1d}. Both models with thermal conduction 
lead to formation of a $\sim 4$-kpc long temperature precursor, and the 
difference between these models and the hydrodynamic model is much 
more prominent than that seen in the surface brightness maps.

Finally, it is possible to measure variations of electron pressure 
in cluster cores by employing the Sunyaev-Zeldovich (SZ) effect \citep{SZ1972}, 
which manifests itself as distortions of the cosmic microwave background (CMB) radiation 
in the direction of galaxy clusters due to Compton scattering of the CMB photons 
off the hot ICM electrons. The magnitude of the SZ effect is proportional to the 
integral of electron pressure along the line of sight. Therefore, we can produce
the SZ map divided by the background model from our simulations: 
\beq
p_{e,{\rm proj}}(x,y) = \frac{ \int_0^{\infty}  (\nesim \Tsim)  \times  (\nez  T_{e0}) \did z }
						 {  \int_0^{\infty} \nez  T_{e0}  \did z}.
\label{eq:sz}
\eeq
The result is shown in the lower panel of \figref{fig:tproj1d}: as one may have 
expected, the conductive precursor is now less prominent than in the projected 
temperature, but stronger than in the surface brightness profiles.

\subsection{Anisotropic thermal conduction}

\begin{figure*}
\centering
\includegraphics[width=1.9\columnwidth]{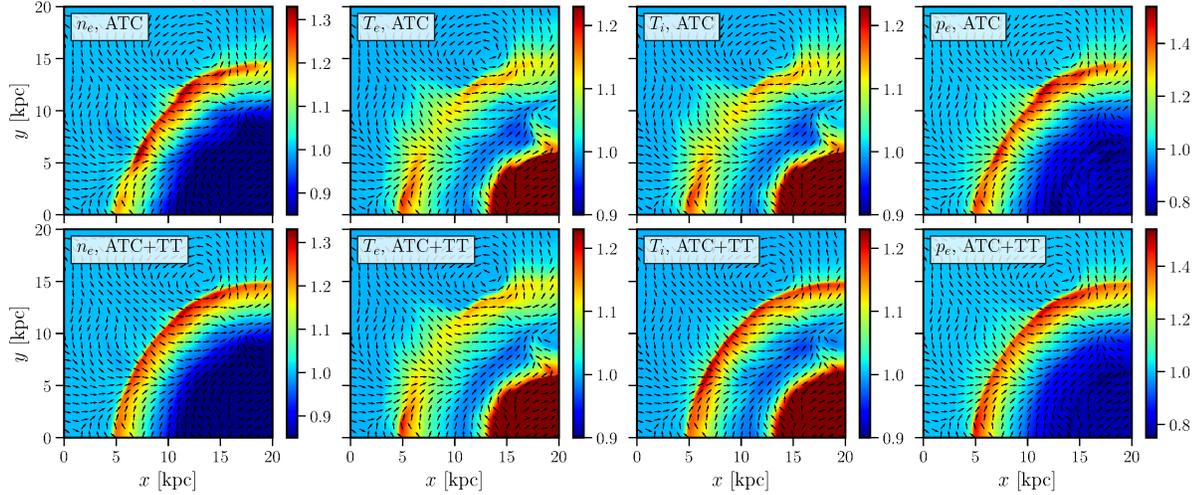}
\caption{ Midplane (z=0) 3D slices of the electron density (left column), electron temperature 
(second from left), ion temperature (second from right), and electron pressure 
(right) for the model with thermal conduction only (upper row) and both conduction and 
ion-electron temperature equilibration (lower row). All the quantities are normalized 
to the pre-shock values. The magnetic-field vectors are superimposed. }
\label{fig:slices}
\end{figure*}

Let us now switch to the 3D runs with anisotropic thermal conduction. 
For these simulations, we initialize a weak random isotropic magnetic field with coherence
length $l_B\approx10$ kpc (as justified in Section~\ref{sec:setup}). 
Anisotropic 
conduction imprints the characteristic correlation scale of the magnetic field 
onto the density and temperature structure of the shock front, as seen clearly in 
\figref{fig:slices}. When ion-electron equilibration is instant, both temperature 
and density jumps of the shock become distorted. The distortion is caused 
by two effects. First, the jumps are smoothed in regions where the magnetic field 
is largely radial, while they stay sharper where the field is close to tangential. 
Second, the shock front becomes corrugated because strong thermal conduction 
along radial field lines reduces the temperature jump (as seen in \figref{fig:1dsim} 
for the 1D models), the shock becomes more isothermal and slows down, while 
if the field is tangential, it propagates at the adiabatic speed of sound as in ideal 
hydrodynamics. 
A finite temperature equilibration 
time attenuates the link between the ions and electrons, allowing the former 
to retain the sharpness of the density jump. The shock front is also less corrugated 
because the ions now preserve their temperature jump. The electron temperature map, however, 
stays qualitatively unchanged. 

\begin{figure*}
\centering
\includegraphics[width=1.8\columnwidth]{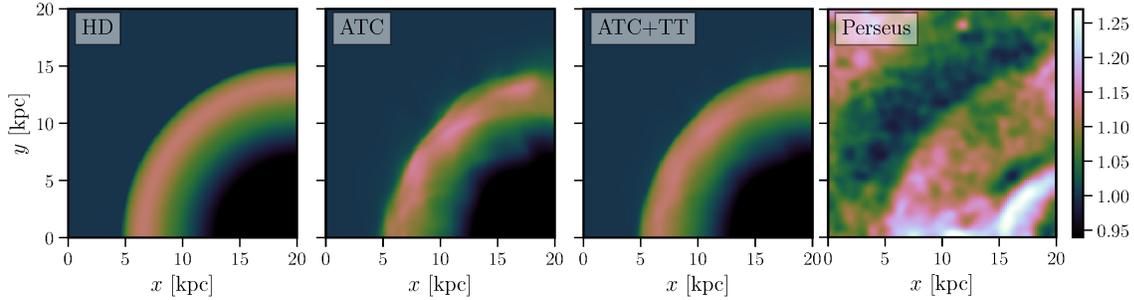}
\caption{ The mock surface brightness maps generated from the 3D simulations 
of weak shocks using \eqref{eq:SB} (the models are the same as 
in \figref{fig:slices} with the addition of pure hydrodynamics in the left panel) 
compared with a {\it Chandra} X-ray image of the weak shock 
in the Perseus cluster (same as in \figref{fig:pers_sect}). 
The images are normalized to the surface brightness in the pre-shock region. }
\label{fig:sb2d}
\end{figure*}

We now use the 3D runs to produce surface brightness maps using the same approach 
as before (equation~\ref{eq:SB}). The results are demonstrated in \figref{fig:sb2d}. 
The model with instant ion-electron equilibration expectedly leads to noticeable 
distortions of the shock front, even though projection effects blur the density 
variations seen more prominently in \figref{fig:slices}. Introducing gradual 
ion-electron equilibration yields a surface brightness map hardly distinguishable 
from pure hydrodynamics. The current Perseus data (right panel of 
\figref{fig:sb2d}) does not allow one to distinguish between the two models. 

\begin{figure*}
\centering
\includegraphics[width=1.3\columnwidth]{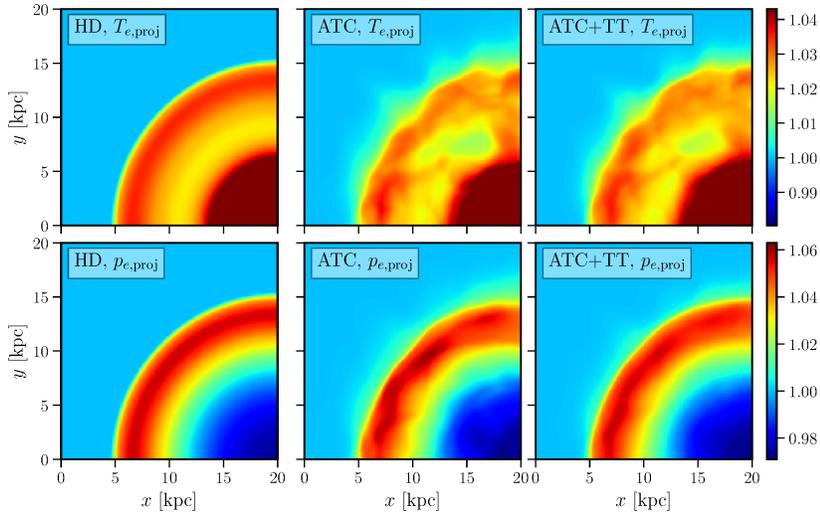}
\caption{Projected temperature and pressure maps calculated from the 3D runs for the same 
				models as in \figref{fig:sb2d}. Both quantities are normalized to the pre-shock region. }  
\label{fig:tproj2d}
\end{figure*}

Analogously, we obtain projected temperature maps by applying equation~\ref{eq:tproj} 
to the 3D simulated temperature. The result can be seen in the upper row of \figref{fig:tproj2d}. Distortions 
of the shock front are now seen in both models because electron temperature is 
relatively unaffected by a finite equilibration time. The maps of electron pressure 
integrated along the line of sight are demonstrated in the lower row of \figref{fig:tproj2d}.

The above results stay qualitatively the same as long as the coherence length of the magnetic field 
is appreciably larger than the shock width ($\sim5$ kpc). If the field is tangled on smaller scales (which, however, 
appears unlikely because the field is naturally expected to be produced by turbulent gas motions 
driven by the jet-bubble dynamics on scales $\sim10$ kpc), the distortions of the shock front 
associated with the topology of the field become averaged out (even more so by the projection effects). 
In this case, one can introduce an 
effective isotropic thermal conductivity as a fraction of the Spitzer value determined by the statistics of the field lines. 
Due to the current lack of the complete theory of high-$\beta$ MHD turbulence, the 
exact value of such effective conductivity cannot be calculated. One way to provide its estimate is to 
apply the theory of strong MHD turbulence, based on the assumption of the so-called 
critical balance \citep{GS1995}, locally to calculate how quickly magnetic field lines diverge and, 
thus, how quickly thermal electrons spread in space \citep{Narayan2001}. Such calculation 
leads to effective conductivities of order $1/5$--$1/3$ of the Spitzer value. Then, the case of 
small-scale magnetic fields is simply reduced to a 1D model with a smaller conductivity. 
The opposite case of a very large-scale magnetic field on scales of tens of kpc is trivial and 
described by the 1D model at full Spitzer conductivity along the direction of the field, 
and ideal hydrodynamics (assuming the field is weak) in the perpendicular plane.

\section{Discussion}
\label{sec:disc}

\subsection{Relevance for present and future observations}

The 900\,ks {\it Chandra}/ACIS imaging spectroscopy of weak shocks in the core of the Perseus cluster currently give us our best view of bulk transport properties on scales approaching the electron mean-free-path.  Even this dataset, however, is not able to provide tight constraints on the presence or absence of conductive precursors once finite electron-ion equilibration is taken into account.  Significantly deeper {\it Chandra}/ACIS observations (bringing the core exposure to $\sim3$\,Ms) would permit sufficiently accurate measurements of (projected) temperature profiles across the shock to address suppression of electron thermal conduction parallel to the magnetic field, providing an important complement to existing limits on perpendicular heat transport at cold fronts.

The equilibration issue can be sidestepped in our study of electron heat transport if we are able to measure electron and ion temperatures separately. This requires high-spectral resolution so that the (electron) bremsstrahlung can be cleanly separated from the ionic emission lines, and it requires high signal-to-noise so that the ion temperature can be determined from the ratio of charge states of given elements. With only modest spectral resolution ($E/\Delta E\sim100$), {\it Chandra}/ACIS is unable to conduct such a study.  Separate measurements of $T_e$ and $T_i$ were performed in the Perseus cluster by the micro-calorimeter on {\it Hitomi} satellite \citep{hitomi}, but the limited spatial resolution of these measurements meant that the spectrum encompassed essentially the entire inner core of the cluster.  However, ATHENA (c2031, \citealt{Athena}) and {\it Lynx} (c2040, \citealt{lynx}) will enable spatially resolved microcalorimeter resolution ($E/\Delta E\sim1000$) spectroscopy, allowing us to map $T_e$ and $T_i$ separately.  {\it Lynx}, with its subarcsecond spatial resolution, will be a particularly powerful observatory for probing the physics of transport on mean-free-path scales in ICM plasmas.

\subsection{Applicability to weak shocks in clusters other than Perseus}

In our study, we have used the weak shock in the Perseus cluster as a fiducial model. 
We have done so for several reasons. First, the shock appears remarkably regular, 
without any significant distortions of the front, which allows one to put constraints 
on processes that can potentially deform the shock. Second, the {\it Chandra} X-ray 
observatory has provided a wealth of high-resolution data for the Perseus core, allowing us to limit 
the shock width down to $\sim1$ kpc. Finally, because the shock is very weak ($M\sim1.2$), 
nonlinear kinetic effects could be subdominant (but see Section~\ref{sec:coll}), and 
it should be approximated by fluid models better. However, there is at least one more example 
of a prominent weak shock in a cluster cool core: the shock in the giant elliptical galaxy 
M87, located $\approx 13$ kpc from the cluster center, exhibits very similar qualities 
(regularity, low Mach number, and even the electron mean free path $\sim0.1$ kpc), however the 
signal-to-noise ratio of its observations by {\rm Chandra} is significantly lower than for Perseus. 
All our results should apply to M87, as well as any very weak shock in cluster 
cores where the electron mean free path is such that the size 
of the potential conductive precursor ($\sim40$ mean free paths) is significant 
to distinguish between the different models.

\subsection{Collisional vs. collisionless shocks}

\label{sec:coll}

We have studied the propagation of a weak shock with Mach number $M\sim1.2$ in the ICM in a 
number of fluid models. In the Perseus cluster, the electron mean free path in the vicinity 
of the shock is $\sim0.1$ kpc, which is 4 times smaller that the resolution of {\it Chandra}. 
Therefore we cannot infer from the data whether the shock is collisional or collisionless. 
Nevertheless, there are several {\it Chandra} observations of merger shocks at higher Mach numbers $M\sim2$--3 
on cluster outskirts where the mean free path reaches $\sim20$ kpc. These observations indicate 
the presence of collisionless processes at play: e.g., \cite{Mark2006} argued that the 
electron-ion equilibration time at the shock in the Bullet cluster is much shorter than the 
Spitzer value, while \cite{Russell2012} were able to measure the width of one of the two 
shocks in Abell~2146 at $\sim6$ kpc, roughly 4 times smaller than the Coulomb mean free path.

Let us discuss the latter observation first. 
We note that even if the shock is collisionless, i.e., there is ion scattering by plasma instabilities 
that generates entropy at the shock, the electrons might still be able to transfer heat across 
the shock, even though they are compressed with the ions by the electric fields ensuring 
plasma quasineutrality, unless electron plasma instabilities develop as well. In this case, 
when the heat flux is still mediated by Coulomb interactions, a 
conductive precursor can develop in the upstream region 
identically to our simulations. However, this may not be the case, at least for higher Mach number $M\sim2$--3 shocks, as 
we speculate below.

\cite{Mark2006} showed that the electron temperature behind the shock in the Bullet cluster 
quickly becomes close to the ion post-shock temperature, suggesting the presence of a kinetic 
mechanism that heats the electrons above the adiabatic compression. \cite{Mascolo}, however, 
used observations of the SZ effect by the Atacama Large Millimeter/submillimeter Array and Atacama Compact Array 
to argue that collisional electron-ion equilibration leads to a better agreement between the radio and {\it Chandra} X-ray data. 
A general discussion on the electron-ion temperature ratio at collisionless shocks can be found in \cite{Vink}. 
Fundamentally, this process needs to be 
studied by means of particle-in-cell plasma simulations. However, the regime of low Mach numbers 
and high plasma beta relevant for the ICM was investigated only recently \citep{Guo2014a,Guo2014b,Guo2017,Guo2018,
Ha2018,Kang2019}. In particular, \cite{Guo2017} proposed a mechanism for electron heating at quasi-perpendicular 
shocks via betatron acceleration of electrons by the amplified magnetic fields (both due to the shock itself 
and the associated ion instabilities in the downstream region) combined with scattering by electron whistler 
waves that disrupts the conservation of electron magnetic moment, thermalizes particles, and generates entropy. 
For our problem, the heating itself is irrelevant, because for very weak shocks the difference between 
the shock and adiabatic heating is small. On the other hand, the potential generation of whistlers at the 
shock is important, because even for small shock compression ratios of $\sim1.3$ the electron temperature anisotropy 
generated by the amplified field via adiabatic invariance may exceed the whistler instability threshold:
\beq
\frac{T_{\perp}}{T_{\parallel}}-1 \eqsim \frac{0.21}{\beta_{\parallel}^{0.6}}, 
\label{eq:whist}
\eeq 
where $T_{\perp}$ and $T_{\parallel}$ are the perpendicular and parallel (to the local magnetic field) 
temperatures, and $\beta_{\parallel}$ is the ratio of the parallel electron pressure to the magnetic-energy 
density \citep{Gary2005}. From adiabatic invariance, the temperature ratio is roughly the 
amplification factor of the magnetic field (equal to the shock compression ratio), which is 
$\sim1.3$ for Mach number $M\sim1.2$, so the left-hand side of \eqref{eq:whist} is $\sim0.3$. 
The electron plasma beta in cluster cores is typically $\sim100$, which gives the right-hand 
side of \eqref{eq:whist} $\sim0.01$. We see that the instability threshold is clearly exceeded even 
for very weak shocks. In case the whistlers are indeed important, our results can be reformulated 
as follows: if no precursors or deformations of the shock are observed, this could be a 
signature of electron scattering by whistler waves at the shock.

\section{Conclusion}
\label{sec:concl}

In this work, we have studied different models of propagation of a very weak shock in 
the ICM, using the weak shock with $M\approx 1.2$ observed in the core of 
the Perseus cluster as a fiducial model. Compared to ideal hydrodynamics, introducing 
isotropic thermal conduction at full Spitzer value leads to formation of a conductive 
precursor with length of order 40 electron mean free paths ($\sim4$ kpc) and a 
decrease of the temperature jump at the shock. When the thermal link between 
the electrons and ions is reduced by a finite equilibration time, the density precursor 
is significantly suppressed, while the electron temperature profile stays roughly the same. 
There is no conclusive evidence of a precursor in the X-ray images of the Perseus 
weak shock we have analyzed. %%A more prominent difference from hydrodynamics is seen in the maps of projected temperature and electron pressure, where the precursor is preserved.

Anisotropic thermal conduction across the weak shock produces temperature and 
density distortions along the shock that are coherent with the structure of the 
magnetic field, mainly its parallel correlation length. The electron temperature 
distortions are preserved almost intact in the two-temperature model, while 
the density variations are largely wiped out and appear close to ideal hydrodynamics.

\section*{Acknowledgements}

C.S.R. thanks the UK Science and Technology Facilities Council (STFC) for support under the New Applicant grant ST/R000867/1, and the European Research Council (ERC) for support under the European Union’s Horizon 2020 research and innovation programme (grant 834203). EC acknowledges support by the Russian Science Foundation grant 19-12-00369.

%%%%%%%%%%%%%%%%%%%%%%%%%%%%%%%%%%%%%%%%%%%%%%%%%%%%

\bibliographystyle{mn2e}
\bibliography{bibliography}

\section*{Data Availability}

The data underlying this article will be shared on reasonable request to the corresponding author.

\end{document}